# FINDING BEAM LOSS LOCATIONS AT PIP2IT ACCELERATOR WITH OSCILLATING DIPOLE CORRECTORS*

A. Shemyakin[†], Fermilab, Batavia, IL 60510, USA

*Abstract*

The PIP2IT accelerator was assembled in multiple stages in 2014 – 2021 to test concepts and components of the future PIP-II linac that is being constructed at Fermilab. In its final configuration, PIP2IT accelerated a 0.55 ms x 20 Hz x 2 mA H- beam to 16 MeV. To determine location of the beam loss in the accelerator's low-energy part, where radiation monitors are ineffective, a method using oscillating trajectories was implemented. If the beam is scraped at an aperture limitation, moving its centroid with two dipole correctors located upstream and oscillating in sync, produces a line at the corresponding frequency in spectra of BPM sum signals downstream of the loss point. Comparison of these responses along the beam line allows to find the loss location. The paper describes the method and results of its implementation at PIP2IT.

## INTRODUCTION

The PIP-II Injector Test (PIP2IT) [1, 2] was an H- ion linac modelling the front end of the PIP-II accelerator currently under construction at Fermilab [3]. In its final config-uration, the PIP2IT consisted of a 30 kV, 15 mA H- DC ion source, a 2 m long Low Energy Beam Transport (LEBT), a 2.1 MeV CW 162.5 MHz RFQ, a 10 m Medium Energy Beam Transport (MEBT), two cryomodules (HWR and SSR1) accelerating the beam up to 16 MeV, a High Energy Beam Transport (HEBT), and a beam dump (Fig. 1).

Beam loss inside the cryomodules was measured by comparison of the beam current read by beam current monitors (ACCT) placed at the exits of the MEBT and SSR1. This comparison indicated the beam loss in the long-pulse mode ~2%. However, such measurement could not point out to a specific location where the loss occurred. In the last days of PIP2IT run, a different method was implemented, where the Beam Position Monitor (BPM) signals were used to identify the loss location.

## METHOD

The method is a development of the idea originally proposed by V. Lebedev and used in CEBAF [4]. It relies on the usually sharp dependence of the current loss on the beam position at the location of the loss. In such case, oscillating a dipole corrector current upstream of the loss location produces a signal at that frequency in BPM sum signals (intensities) downstream. Such measurement does not provide an absolute value of the loss but rather the difference in the loss over the range of the beam oscillation. While this loss variation can be low, the detection at a fix frequency greatly improves the overall sensitivity. For sufficiently long measurement time, even oscillations with amplitude small enough to do not affect the beam emittance can result in a detectable signal.

Ref. [5] proposed to oscillate simultaneously two correctors (in one plane) to check in one measurement all locations in a beam line or linac. The initial test of the procedure is described in Ref. [6]. The proposal is to oscillate currents in two dipole correctors separated by the betatron phase advance of $\varphi_x \neq \pi n$ with a specific choice of amplitudes of resulting deflections $\theta_1$ and $\theta_2$ and the time phase difference $\varphi_t$ (similar to Ref. [7]):

$$\theta_2\sqrt{\beta_{x2}} = \theta_1\sqrt{\beta_{x1}}, \quad \varphi_t = \pi + \varphi_x, \quad (1)$$

where $\beta_{x1}$ and $\beta_{x2}$ are betatron functions in the location of corresponding corrector. At these conditions, the deviation of the trajectory downstream is simplified to

$$x_0(z,t) = \theta_1\sqrt{\beta_x(z)\beta_{x1}} \sin\varphi_x \sin(\omega t + \varphi_1(z)), \quad (2)$$

where $\beta_x(z)$ is the beta-function along the line. The Fourier component of BPM readings at the oscillation frequency $\omega/2\pi$ is determined by the beta-function in the BPM location, and its phase relates to the betatron phase advance $\varphi(z)$ as $\varphi_1(z) = \varphi(z) + \varphi_x$. Oscillation described by Eq. (2) move the beam around a circle in canonical phase coordinates, shifting the beam by the same portion of its rms size $\sigma_b = \sqrt{\beta_x(z)\varepsilon_0}$ everywhere along the beam line ($\varepsilon_0$ is the rms beam emittance).

Let's assume that the 1D current density distribution is scaled in various locations as the beam rms size:

$$j(x) = \frac{I_0}{\sigma_b} J\left(\frac{x}{\sigma_b}\right), \quad (3)$$

where $I_0$ is the total beam current, and $J\left(\frac{x}{\sigma_b}\right)$ is a dimensionless function, the same for all locations. If a flat scraper is inserted into the beam to the distance *d* from the beam center, the intercepted current $I_s$ is modulated at the oscillation frequency:

$$I_s = \int_d^\infty j(x-x_0)dx = \int_d^\infty j(x)dx + j(d)x_0 - \frac{dj}{dx}(d) \cdot \frac{x_0^2}{2} + \cdots \approx \int_d^\infty j(x)dx + j(d)A_s \sin(\omega t + \varphi_1(z_s)) + j'(d) \cdot \frac{1}{2}(A_s \sin(\omega t + \varphi_1(z)))^2 \equiv I_{s0} + I_{s1}\sin(\omega t + \varphi_1(z_s)) + I_{s2}(1 - \cos(2 \cdot (\omega t + \varphi_1(z_s)))). \quad (4)$$

where $A_s \equiv \theta_1\sqrt{\beta_x(z_s)\beta_{x1}}$ is the trajectory oscillation amplitude at the scraper location $z_s$. The amplitude of the first harmonic depends only on the relative penetration of the scraper:

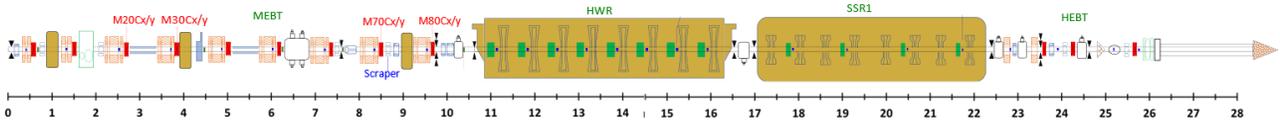

Figure 1: Schematic of the PIP2IT downstream of the RFQ. The scale is in meters. Dipole correctors are represented by red rectangles, and BPMs are by blue dots. Courtesy of L. Prost.

$$\frac{I_{s1}}{I_0} = \frac{\theta_1 \sqrt{\beta_{x1}}}{\sqrt{\varepsilon_0}} J\left(\frac{d}{\sigma_b}\right). \quad (5)$$

The phase of the beam loss oscillations is determined by the betatron phase in the location of scraper $\varphi_1(z_s)$, independently on where the current loss is measured. Note that if the scraper is inserted from the negative side, the phase reported in Fourier analysis is shifted by $\pi$.

The term $I_{s2}$ in Eq. (4) describes nonlinearity of the beam loss and appears in the Fourier spectrum as the second harmonic. In the measurements described below, it was always found below the noise level, and all detectable signals were dominated by the linear component. In part, it means that the oscillations did not increase the average loss level.

The errors of the values measured with oscillations (both positions and current loss) can be estimated assuming that the rms noise at the oscillation frequency is the same as at other frequency components in the measured spectrum, and the noise phase is random [8]:

$$\sigma_a = \sqrt{\frac{\overline{a_k^2}}{2}}, \quad \sigma_\varphi = \frac{\sigma_a}{a}, \quad (6)$$

where $\sigma_a$ is the rms error of measuring an amplitude $a$, $\overline{a_k^2}$ is the average value of all squared Fourier amplitudes (excluding the driving frequencies and their second harmonics), and $\sigma_\varphi$ is the rms error of the measured phase.

## MEASUREMENTS

The measurements to find the loss location were performed using the pairs of dipole correctors in the MEBT shown in Fig. 1 as (M20C, M30C, upstream) or (M70C, M80C, downstream). Each pair (either horizontal, X, or vertical, Y) was oscillated in sync with amplitudes and the phase offset calculated with Eq. (1) using the MEBT optics functions measured with differential trajectories [9].

The measurements were performed at operational parameters of PIP2IT except the pulse length was reduced to 10 µs. The pulse current of the beam coming out of the RFQ was 5 mA. In the MEBT, the bunches were scraped transversely, and half of them was removed by the chopping system, so the beam pulse current at the end of the MEBT was 1.8 mA. Transmission through the cryomodules in the measurements presented in this paper was not optimized and was higher than in long-pulse measurements.

The program recorded data and moved to the next corrector value only when the time stamps for all channels were aligned. Because of difficulties with synchronization of the front ends, only a portion of all pulses was used. While the pulse rate was 20 Hz, the recording frequency was ~3 Hz so that the total time for one measurement varied between 1.5 and 7 min. Effective frequency of the corrector current oscillations was ~0.1 Hz. In the later measurements, two pairs of correctors (X and Y) were oscillated at the same time with different periods to speed up the measurements. Such simultaneous oscillation did not affect the results or error bars.

Three signals were recorded from each BPM, X/Y positions and intensity (sum). A Discrete Fourier Transform (DFT) was applied to all signals in MathCad, providing for each channel the amplitude and oscillation phase. An example of BPM response to oscillation is shown in Fig. 2. Note that this X BPM responds to oscillation in both planes because it is located downstream of a solenoid.

Responses of BPM positions to excitation by different pairs of correctors in the same plane were nearly identical after scaling for difference in the initial betatron functions and phases (Fig. 3). It indicates a good quality of the optical model in MEBT and reproducibility of the measurements.

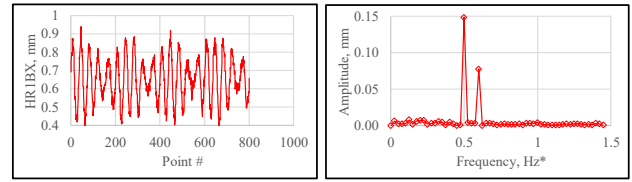

Figure 2: Response of the first HWR X BPM to oscillating of two pairs (X, 40.1 points period, and Y, 33.4 points period) of correctors (M20C, M30C) (left) and the relevant part of its spectrum (right). 802 points. *Horizontal axis on the right plot is as if frequency of recording were 20 Hz.

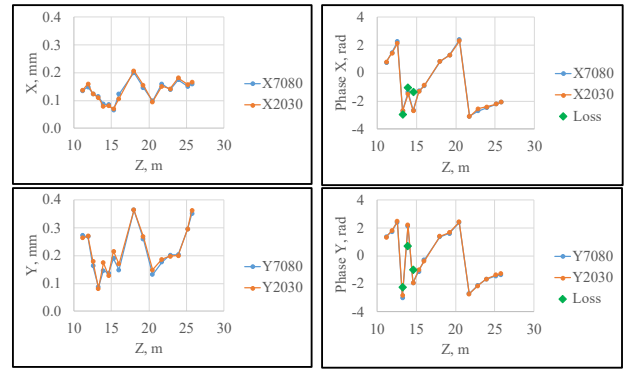

Figure 3: Comparison of oscillation amplitudes (left) and phases (right) of BPM in-plane positions in oscillation of (M20C, M30C) and (M70C, M80C) corrector pairs. The former data are adjusted for the initial deflection amplitude (by 0.9/1.3 in X/Y) and phase offset (-0.1/-0.8 rad in X/Y). "Loss" points show phases of differential intensities.

A typical response of BPM intensities is shown in Fig. 4. The plotted values are ratios of the recorded BPM response amplitude to the average value in the same channel. They show by how much the relative beam loss changes when positions are changed as in Fig. 3. In each location, such value is defined by a vector sum of changes in all losses upstream. For example, the beam is significantly scraped, by design, at the MEBT absorber (primarily in Y direction). In the case of using the upstream correctors, this loss dominates all signals downstream. Also, the curves are not necessarily monotonous. If, let say, in horizontal direction the beam is scraped first on the left side but near the next downstream BPM on the right side, moving the beam to the left will increase the beam loss in the first location but may decrease the overall loss in the second location.

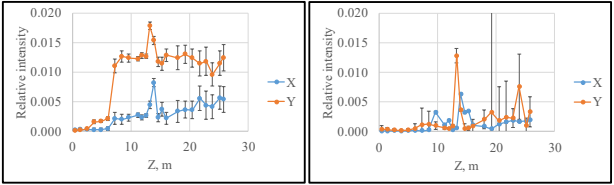

Figure 4: Amplitudes of relative BPM intensity response to oscillations in X and Y pairs of (M20C, M30C) (left) and (M70C, M80C) (right) correctors. 2005 points for the blue curve in the right plot (X7080) and 802 points for others.

A more informative approach is to analyze the differences $J_{i,k}$ between neighboring BPMs, $J_{i,k} = \frac{A_{i,k}}{Int_i} - \frac{A_{i-1,k}}{Int_{i-1}}$, where $A_{i,k}$ and $Int_i$ are the intensity reported by BPM $i$ in the $k$ sample during oscillation and its average value, correspondingly. The DFT amplitude of $\{J_{i,k}\}$ characterizes the local change in the beam loss occurring between two neighbouring BPMs and is unaffected by the loss upstream (assuming that the loss is small and the betatron phase advance between loss locations is far enough from $2\pi n$). Correspondingly, the patterns induced by oscillation of different pairs of correctors become similar, different primarily due to difference in the excitation amplitude and statistical errors (Fig. 5).

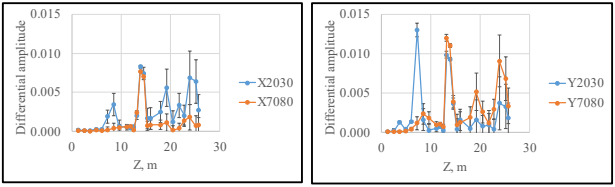

Figure 5: DFT amplitudes of differential BPM signals. Analysis of the same data as in Fig. 4. The curves arranged differently for comparison.

To test the procedure, a 6.4% loss was artificially introduced by inserting a vertical MEBT scraper in the MEBT into the 1.8 mA beam. The resulting change in the differential signal of the BPM right downstream of the scraper (Fig. 6) clearly indicates the loss. Location of the loss can be determined more accurately by comparing the oscillation phases in the signals of BPM positions and differential intensities. These phases are defined by the betatron phase at location of BPM positions and of the loss (in this case, position of the scraper), correspondingly. The "Loss" point in Fig. 6 right has the ordinate equal to the oscillation phase of the difference between intensities of BPMs at Z= 9.6 m and 8.4 m, and its abscissa is the known longitudinal position of the scraper. This comparison predicts the loss location reasonably well, within 0.2 m.

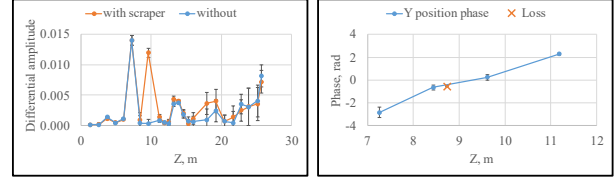

Figure 6: Effect of inserting a scraper on differential losses induced by oscillation of Y2030 correctors. 401 points.

Actual losses in the MEBT, represented by peaks at Z <10 m, also agree well with known aperture limitations or location of intentional scraping at the MEBT absorber (at Z=6.8 m).

However, results in the cryomodules (11.2 m <Z<22.8 m) are more complicated. On one hand, in all sets of data the differential BPM intensity signals are clearly seen (i.e. exceed 3 rms error) in 3 HWR BPMs located at 13.2, 13.9, and 14.6 m, indicating a loss in this portion of the cryomodule. On the other hand, no clear correlation was observed between amplitudes of these signals and the loss reported by comparison of ACCTs. For example, the data sets reported in Fig. 5 and 6 were taken in different days at different settings of the dipole correctors (but at the same beam current and pattern of oscillations). The loss in the cryomodules reported by ACCT comparison was 2.7% for the Fig. 5 case and 6.7% for Fig. 6 case, while the differential BPM intensities moved in opposite direction. Also, their phases fit worse to the BPM position oscillation phases ("Loss" points in Fig. 3) than in the scraper case. While no fully satisfactory explanation was found for this effect, there are indications that it could be related to inaccurate reporting of beam positions and intensities by BPMs when they are irradiated by beam losses or secondary particles. At small incident angles, the secondary electron emission yield can be very large, which might significantly affect the BPM signals.

## SUMMARY

The procedure of finding a beam loss location by oscillating a pair of correctors was successfully tested by inserting a beam scraper into the beam. Applying the method at the cryomodules indicated the loss location in the middle of the HWR cryomodule.

## ACKNOWLEDGEMENTS

Author is thankful to the entire PIP2IT team for assistance with the measurements. The optical functions used for calculation of the amplitudes and phases of oscillating correctors were simulated by A. Saini. All data acquisition programs were written by W. Marsh.

# REFERENCES

[1] P. Derwent *et al.*, "PIP-II Injector Test: Challenges and Status", in *Proc. 28th Linear Accelerator Conf. (LINAC'16)*, East Lansing, MI, USA, Sep. 2016, pp. 641-645.
doi:10.18429/JACoW-LINAC2016-WE1A01

[2] E. Pozdeyev *et al.*, "Beam Commissioning and Integrated Test of the PIP-II Injector Test Facility", presented at LINAC'22, Liverpool, UK, September 2022, paper MO1PA01, this conference

[3] "The Proton Improvement Plan-II (PIP-II) Final Design Report", 2021 (unpublished).

[4] V. Lebedev, private communication, 2016

[5] A. Shemyakin, "Finding beam loss locations in a linac with oscillating dipole correctors", Fermilab Technical Memo FERMILAB-TM-2708-AD, Fermilab, Batavia IL, USA, May 2019; http://arxiv.org/abs/1905.00363

[6] A. V. Shemyakin, R. Prakash, and K. Seiya, "Finding Beam Loss Locations in a Linac with Oscillating Dipole Correctors", in *Proc. NAPAC'19*, Lansing, MI, USA, Sep. 2019, pp. 663-666.
doi:10.18429/JACoW-NAPAC2019-WEPLM02

[7] Xiaobiao Huang, "Linear optics and coupling correction with closed orbit modulation", Phys. Rev. Accel. Beams, vol. 24, p. 072805, 2021.
doi:10.1103/physrevaccelbeams.24.072805

[8] A. Shemyakin, "Orbit response measurements with oscillating trajectories", Fermilab, USA, FERMILAB-TM-2763-AD, 2021, https://arxiv.org/abs/2109.11589

[9] A. V. Shemyakin *et al.*, "Experimental Study of Beam Dynamics in the PIP-II MEBT Prototype", in *Proc. HB'18*, Daejeon, Korea, Jun. 2018, pp. 54-59.
doi:10.18429/JACoW-HB2018-MOP1WB03